\documentclass{aa} 

\usepackage[varg]{txfonts}
\usepackage[utf8]{inputenc}
\usepackage{textcomp}
\usepackage{gensymb}
\usepackage{graphicx}
\usepackage{natbib}
\usepackage[super]{nth}

\bibpunct{(}{)}{;}{a}{}{,} 

\begin{document}

\title{Dynamical modeling validation of parent bodies associated with newly discovered CMN meteor showers}

\author{Damir Šegon\inst{1,\thanks{E-mail: damir@astro.hr}}
\and Jérémie Vaubaillon\inst{2}
\and Peter S. Gural\inst{3}
\and Denis Vida\inst{4,5,6}
\and \\Željko Andreić\inst{7}
\and Korado Korlević\inst{8}
\and Ivica Skokić\inst{4,9}
}
\institute{Astronomical Society Istra Pula, Park Monte Zaro 2, 52100 Pula, Croatia
\and IMCCE, Observatoire de Paris, 77 Avenue Denfert Rochereau, F-75014 Paris, France
\and Gural Software Development, 351 Samantha Drive, Sterling, Virginia, USA 20164
\and Astronomical Society "Anonymus", B. Radić 34, 31550 Valpovo, Croatia
\and Department of Earth Sciences, University of Western Ontario, London, Ontario, N6A 5B7, Canada
\and Department of Physics and Astronomy, University of Western Ontario, London, Ontario, N6A 3K8, Canada
\and Faculty of Mining, Geology, and Petroleum Engineering, University of Zagreb, Pijerottijeva 6, 10000 Zagreb, Croatia
\and Višnjan Science and Education Center, Istarska 5, 51463 Višnjan, Croatia
\and Astronomical Institute of the Czech Academy of Sciences, Fričova 298, 251 65 Ondřejov, Czech Republic
}
\date{Received June 13, 2016 / Accepted October 25, 2016}
\abstract 
{Results from previous searches for new meteor showers in the combined Croatian Meteor Network and SonotaCo meteor databases suggested possible parent bodies for several newly identified showers.} 
{We aim to perform an analysis to validate the connection between the identified showers and candidate parent bodies.} 
{Simulated particles were ejected from candidate parent bodies, a dynamical modeling was performed and the results were compared to the real meteor shower observations.} 
{From the 13 analysed cases, three were found to be connected with comets, four with asteroids which are possibly dormant comets, four were inconclusive or negative, and two need more observational data before any conclusions can be drawn.} 
{}

\titlerunning{Associations of parent bodies and meteor showers from the CMN search}
\authorrunning{Šegon et al.}

\maketitle

\section{Introduction}
When one is attempting to associate newly discovered meteoroid streams to their parent bodies, there are four critical steps that need to be carried out. The first is obviously the stream discovery through the search of databases and past records, which is ideally performed on meteor data comprised of Keplerian orbital elements. The second phase involves the verification of the meteoroid stream using completely independent meteor databases and stream searches as published online and/or reported in the literature. This is to help validate the existence of the stream. The third step explores the identification of candidate parent bodies, such as comets and asteroids, which show similar orbits to the space-time aggregated meteoroid Keplerian elements of the found stream. However, close similarity of the orbits between a meteoroid stream and a potential parent body is not necessarily conclusive proof of association or linkage, since the two object types (parent body and meteoroid) can undergo significantly different orbital evolution as shown by \citet{vaubaillon2006mechanisms}. Thus the most critical fourth step in determining the actual association is to perform dynamic modeling and orbital evolution on a sample of particles ejected from a candidate parent body. Given a comet's or asteroid's best estimated orbit in the past, and following the ejected stream particles through many hundreds to thousands of years, one looks for eventual encounters with the Earth at the time of meteor observation, and whether those encounters have a geometric similarity to the observed meteoroids of the stream under investigation. The work by \citet{mcnaught1999leonid} demonstrates this point. However, this current paper follows the approach of \citet{vaubaillon2005new} in focusing on the results of the dynamical modeling phase and is a culmination of all the steps just outlined and performed on new streams discovered from recent Croatian Meteor Network stream searches. The application of dynamical stream modeling indicates, with a high level of confidence, that seven new streams can be associated to either comets or asteroids, the latter of which are conjectured to be dormant comets.

\section{Processing approach}
The seven streams and their hypothetical parent body associations were initially discovered using a meteor database search technique as described in \citet{segon2014parent}. In summary, the method compared every meteor orbit to every other meteor orbit using the combined Croatian Meteor Network \citep{segon2012croatian, korlevic2013croatian}\footnote{\label{cmn_orbits}\url{http://cmn.rgn.hr/downloads/downloads.html#orbitcat}.} and SonotaCo \citep{sonotaco2009meteor}\footnote{\label{sonotaco_orbits}\url{http://sonotaco.jp/doc/SNM/index.html}.} video meteor orbit databases, looking for clusters and groupings in the five-parameter, Keplerian orbital element space. This was based on the requirement that three D-criteria (\citet{southworth1963statistics}, \citet{drummond1981test}, \citet{jopek2008meteoroid}) were all satisfied within a specified threshold. These groups had their mean orbital elements computed and sorted by number of meteor members. Mean orbital elements where computed by a simple averaging procedure. Working down from the largest sized group, meteors with similar orbits to the group under evaluation were assigned to the group and eliminated from further aggregation. This captured the known streams quickly, removing them from the meteor pool, and eventually found the newly discovered streams. According to International Astronomical Union (IAU) shower nomenclature rules \citep{jenniskens2006iau}, all results for the stream discoveries were first published. In these cases the search results can be found in three papers posted to WGN, The Journal of the International Meteor Organization \citep{andreic2014, gural2014results, segon2014results}.

Next, the literature was scoured for similar stream searches in other independent data sets, such as the CAMS \citep{rudawska2014new, jenniskens2016cams} and the EDMOND \citep{rudawska2014independent} video databases, to determine the validity of the new streams found. The verified new streams were then compared against known cometary and asteroidal orbits, from which a list of candidate parent bodies were compiled based once again on meeting multiple D-criteria for orbital similarity. Each section below describes in greater detail the unique processes and evidence for each stream's candidate association to a parent body. Besides the seven reported shower cases and their hypothetical parent bodies, the possibility of producing a meteor shower has also been investigated for four possible streams with similar orbital parameters to asteroids, 2002 KK3, 2008 UZ94, 2009 CR2, and 2011 YX62, but the results were inconclusive or negative. The remaining possible parent bodies from the search were not investigated due to the fact that those comets do not have orbital elements precise enough to be investigated or are stated to have parabolic orbits.

The dynamical analysis for each object was performed as follows. First, the nominal orbit of the body was retrieved from the JPL HORIZONS ephemeris\footnote{\label{horizonsJPL}\url{http://horizons.jpl.nasa.gov}} for the current time period as well as for each perihelion passage for the past few centuries (typically two to five hundred years). Assuming the object presented cometary-like activity in the past, the meteoroid stream ejection and evolution was simulated and propagated following \citet{vaubaillon2005new}. In detail, the method considers the ejection of meteoroids when the comet is within 3 AU from the Sun. The ejection velocity is computed following \citet{crifo1997dependence}. The ejection velocities typically range from 0 to \textasciitilde100 m/s. Then the evolution of the meteoroids in the solar system is propagated using numerical simulations. The gravitation of all the planets as well as non-gravitational forces (radiation pressure, solar wind, and the Poynting-Robertson effect) are taken into account. More details can be found in \citet{vaubaillon2005new}. When the parent body possessed a long orbital period, the stream was propagated starting from a more distant period in the past few thousand years. The intersection of the stream and the Earth was accumulated over 50 to 100 years, following the method by \citet{jenniskens2008minor}. Such a method provides a general view of the location of the meteoroid stream and give statistically meaningful results. For each meteoroid that is considered as intersecting the Earth, the radiant was computed following the \citet{neslusan1998computer} method (the software was kindly provided by those authors). Finally, the size distribution of particles intercepting the Earth was not considered in this paper, nor was the size of modeled particles compared to the size of observed particles. The size distribution comparison will be the topic of a future paper.

\section{IAU meteor shower \#549 FAN - 49 Andromedids and Comet 2001 W2 Batters
}
The first case to be presented here is that of meteor shower IAU \#542 49 Andromedids. Following the IAU rules, this shower was first reported as part of a paper in WGN, Journal of International Meteor Organization by \citet{andreic2014}.  Independent meteor shower database searches resulted in confirmation of the existence of this shower, namely \citet{rudawska2015independent} and \citet{jenniskens2016cams}.  The radiant position from the Croatian Meteor Network (CMN) search into the SonotaCo and CMN orbit databases was found to be R.A. = 20.9°, Dec. = +46.7°, with a mean geocentric velocity V$_{g}$ = 60.1 km/s near the center of the activity period (solar longitude $\lambda_{0}$ = 114°, 35 orbits). \citet{rudawska2015independent} found the same radiant to be at R.A. = 19.0°, Dec. = +45.3° and V$_{g}$ = 59.8 km/s ($\lambda_{0}$ = 112.5°, 226 orbits), while \citet{jenniskens2016cams} give R.A. = 20.5°, Dec. = +46.6°, and V$_{g}$ = 60.2 km/s ($\lambda_{0}$ = 112°, 76 orbits). This shower was accepted as an established shower during the 2015 IAU Assembly\footnote{\label{IAU2015}\url{https://astronomy2015.org}.} and is now listed in the IAU meteor database.

At the time of the initial finding, there were 35 meteors associated with this shower resulting in orbital parameters similar to published values for a known comet, namely 2001 W2 Batters. This Halley type comet with an orbital period of 75.9 years, has been well observed and its orbital parameters have been determined with higher precision than many other comets of this type. The mean meteoroid orbital parameters, as found by the above mentioned procedure, are compared with the orbit of 2001 W2 Batters in Table \ref{tab:table1}. Despite the fact that the orbital parameters' distance according to the Southworth-Hawkins D-criteria D$_{SH}$ = 0.14 seems a bit high to claim an association, the authors pointed out the necessity of using dynamic stream modeling to confirm or deny the association hypothesis because of the nearly identical ascending node values. Moreover, changes in 2001 W2 Batters' orbital parameters as far back as 3000 BC, as extracted from HORIZONS, has shown that the comet approached closer to Earth's orbit in 3000 BC than it has during the last few hundred years. Thus stream particles ejected from the comet farther in the past could have the possibility of producing a meteoroid stream observed at the Earth in the current epoch.

\begin{table*}[t] 
        \caption{Orbital parameters for the 49 Andromedids and Comet 2001 W2 Batters with corresponding D$_{SH}$ values. If the value of 112° for the ascending node (from \citet{jenniskens2016cams}) is used instead of the mean value (118°), then the resulting D$_{SH}$ is 0.16. Orbital elements (mean values for shower data): q = perihelion distance, e = eccentricity, i = inclination, Node = Node, $\omega$ = argument of perihelion, D$_{SH}$ = Southworth and Hawking D-criterion with respect to 2001 W2 Batters.}
        \label{tab:table1} 
        \centering 
        \begin{tabular}{l c c c c c c} 
        \hline\hline 
        49 Andromedids & q & e & i & Node & $\omega$ & D$_{SH}$ \\
        References & (AU) & & (\degr) & (\degr) & (\degr) & \\
        \hline 
        1 &  0.918 & 0.925 & 118.2 & 114.0 & 143.1 & 0.14\\
        2 & 0.907 & 0.878 & 119.2 & 112.5 & 142.2 & 0.17\\
        3 & 0.898 & 0.922 & 117.9 & 118.0 & 139.8 & 0.19\\
        2001 W2 Batters & 1.051 & 0.941 & 115.9 & 113.4 & 142.1 & 0\\
        \hline 
        \end{tabular}
                \tablebib{(1) \citet{andreic2014}; (2) \citet{rudawska2015independent}; (3) \citet{jenniskens2016cams}.
        }
\end{table*}

The dynamical modeling for the hypothetical parent body 2001 W2 Batters was performed following \citet{vaubaillon2005new} and \citet{jenniskens2008minor}. In summary, the dynamical evolution of the parent body is considered over a few hundred to a few thousand years. At a specific chosen time in the past, the creation of a meteoroid stream is simulated and its evolution is followed forward in time until the present day. The intersection of the particles with the Earth is recorded and the radiant of each particle is computed and compared to observations. The first perihelion passages were initially limited to 500 years back in time. No direct hits to the Earth were found from meteoroids ejected during the aforementioned period. However, the authors were convinced that such close similarity of orbits may result in more favorable results if the dynamical modeling was repeated for perihelion passages back to 3000 BC. The new run did provide positive results, with direct hits to the Earth predicted at R.A. = 19.1°, Dec. = +46.9°, and V$_{g}$ = 60.2 km/s, at a solar longitude of $\lambda_{0}$ = 113.2°. A summary of the observed and modeled results is given in Table \ref{tab:table2}.

\begin{table}[h]
        \caption{Observed and modeled radiant positions for the 49 Andromedids and comet Batters' meteoroids ejected 3000 years ago.}
        \label{tab:table2} 
        \centering 
        \begin{tabular}{l c c c c} 
        \hline\hline 
        49 Andromedids & R.A. & Dec. & V$_{g}$ & $\lambda_{0}$\\
        References & (\degr) & (\degr) & (km/s) & (\degr)\\
        \hline 
        1 & 20.9 & 46.7 & 60.1 & 114.0\\
        2 & 19.0 & 45.3 & 59.8 & 112.5\\
        3 & 20.5 & 46.6 & 60.2 & 112.0\\
        2001 W2 Batters \\meteoroids, this work & 19.1 & 46.9 & 60.2 & 113.2\\
        \hline 
        \end{tabular}
        \tablebib{(1) \citet{andreic2014}; (2) \citet{rudawska2015independent}; (3) \citet{jenniskens2016cams}.
        }
\end{table}

The maximum difference between the average observed radiant positions and modeled mean positions is  less than 2° in both right ascension and declination, while there are also single meteors very close to the predicted positions according to the model. Since the observed radiant position fits very well with the predictions, we may conclude that there is a strong possibility that comet 2001 W2 Batters is indeed the parent body of the 49 Andromedids shower. The high radiant dispersion seen in the observations can be accounted for by 1) less precise observations in some of the reported results, and 2) the 3000 year old nature of the stream which produces a more dispersed trail. The next closest possible association was with comet 1952 H1 Mrkos but with D$_{SH}$ of 0.28, it was considered too distant to be connected with the 49 Andromedids stream.

\begin{figure}[h]
        \resizebox{\hsize}{!}{\includegraphics{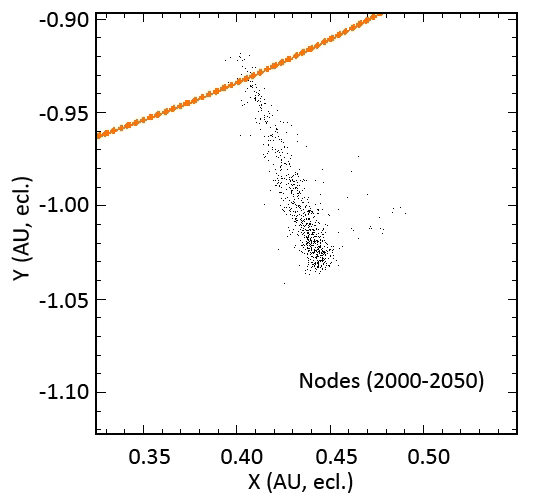}}
        \caption{Location of the nodes of the particles released by 2001 W2 Batters over several centuries, concatenated over the years 2000 to 2050. The Earth crosses the stream.}
        \label{fig:figure1}
\end{figure}

\begin{figure}[h]
        \resizebox{\hsize}{!}{\includegraphics{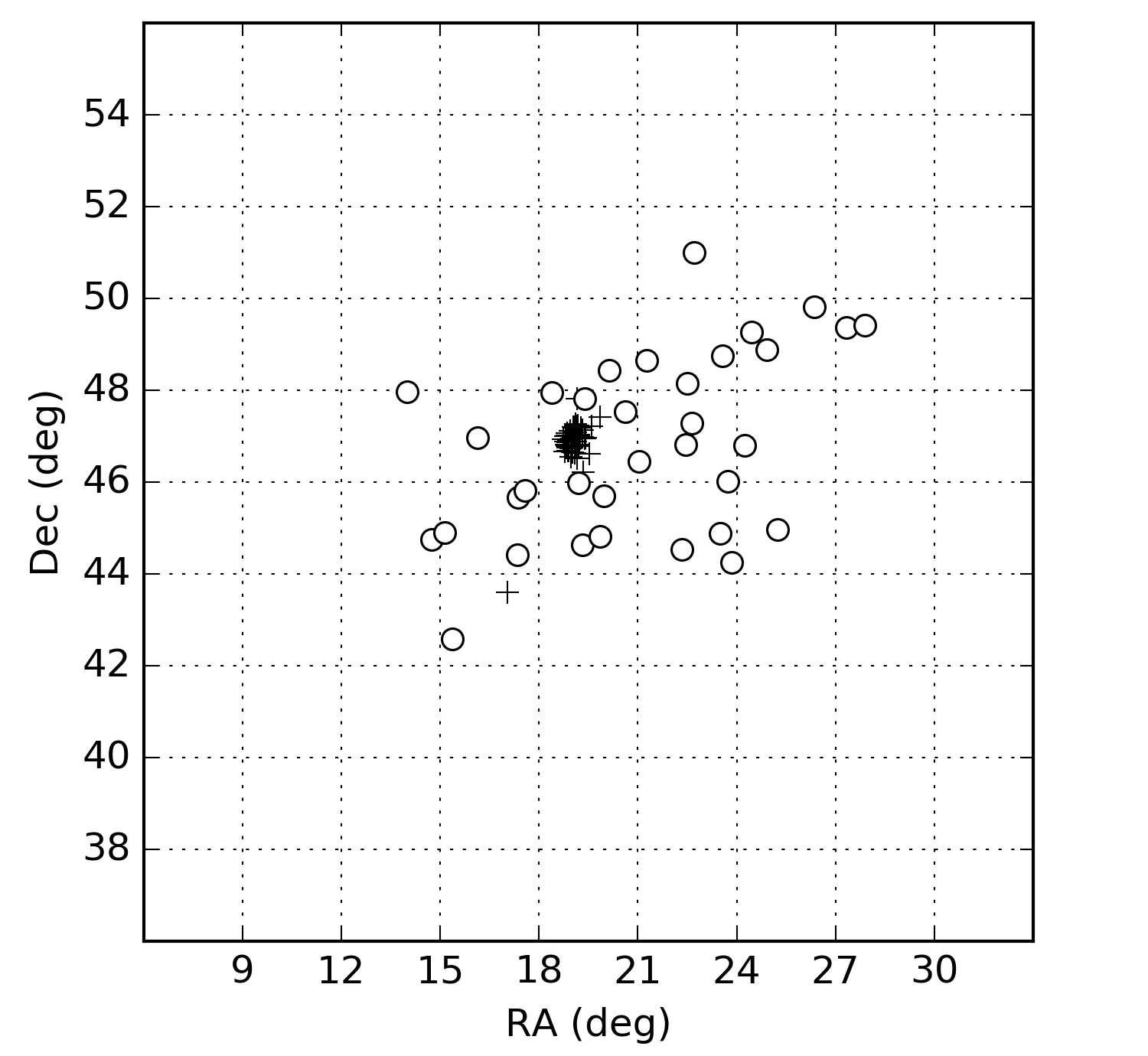}}
        \caption{Theoretical radiant of the particles released by 2001 W2 Batters which were closest to the Earth. The range of solar longitudes for modeled radiants is from 113.0\degr to 113.9\degr. Pluses represent the modeled radiants in the given solar longitude range, while the circles represent the observed radiants during the whole activity of the shower.}
        \label{fig:figure2}
\end{figure}

Figures \ref{fig:figure1} and \ref{fig:figure2} show the location of the stream with respect to the Earth's path, as well as the theoretical radiant. These results were obtained by concatenating the locations of the particles intersecting the Earth over 50 years in order to clearly show the location of the stream (otherwise there are too few particles to cross the Earth each year). As a consequence, it is expected that the level of activity of this shower would not change much from year to year.

\section{IAU meteor shower  \#533 JXA - July $\xi$ Arietids  and comet 1964 N1 Ikeya}
The discovery of the possible meteor shower July $\xi$ Arietids was first published in \citet{segon2014new}. The shower had been found as a grouping of 61 meteoroid orbits, active from July 4 to August 12, peaking around July 21. Three other searches for meteor showers in different meteoroid orbit databases done by \citet{rudawska2015independent}, \citet{jenniskens2016cams}, and \citet{kornovs2014confirmation} found this shower as well, but with slight differences in the period of activity. This shower had been accepted as an established shower during the 2015 IAU Assembly held on Hawaii and is now referred to as shower \#533.

Among the possible parent bodies known at the time of this shower's discovery, comet C/1964 N1 Ikeya was found to have similar orbital parameters as those of the July $\xi$ Arietids. Comet C/1964 N1 Ikeya is a long period comet, having an orbital period of 391 years and contrary to comet 2001 W2 Batters, has less precision in its orbit estimation. A summary of the mean orbital parameters of the shower compared with C/1964 N1 Ikeya are shown in Table \ref{tab:table3}, from which it can be seen that the distance estimated from D$_{SH}$ suggests a possible connection between the shower and the comet. 

\begin{table*}[t]
        \caption{Orbital parameters for the July $\xi$ Arietids and Comet 1964 N1 Ikeya with corresponding D$_{SH}$ values. Orbital elements (mean values for shower data): q = perihelion distance, e = eccentricity, i = inclination, Node = Node, $\omega$ = argument of perihelion, D$_{SH}$ = Southworth and Hawking D-criterion with respect to 1964 N1 Ikeya.}
        \label{tab:table3} 
        \centering 
        \begin{tabular}{l c c c c c c} 
        \hline\hline 
        July $\xi$ Arietids & q & e & i & Node & $\omega$ & D$_{SH}$\\
        References & (AU) & & (\degr) & (\degr) & (\degr) & \\
        \hline 
        1 & 0.883 & 0.965 & 171.6 & 299.0 & 318.0 & 0.10\\
        2 & 0.863 & 0.939 & 171.8 & 292.6 & 313.8 & 0.08\\
        3 & 0.836 & 0.919 & 171.5 & 291.1 & 309.8 & 0.09\\
        4 & 0.860 & 0.969 & 170.4 & 292.7 & 312.4 & 0.08\\
        C/1964 N1 Ikeya & 0.822 & 0.985 & 171.9 & 269.9 & 290.8 & 0\\
        \hline 
        \end{tabular}
        \tablebib{(1) \citet{segon2014new}; (2) \citet{kornovs2014confirmation}; (3) \citet{rudawska2015independent}; (4) \citet{jenniskens2016cams}.
        }
\end{table*}

Similar to the previous case, the dynamical modeling was performed for perihelion passages starting from 5000 BC onwards. Only two direct hits were found from the complete analysis, but those two hits confirm that there is a high possibility that comet C/1964 N1 Ikeya is indeed the parent body of the July $\xi$ Arietids. The mean radiant positions for those two modeled meteoroids as well as the mean radiant positions found by other searches are presented in Table \ref{tab:table4}. As can be seen from Table \ref{tab:table4}, the difference in radiant position between the model and the observations appears to be very significant. 

\begin{table}[h]
        \caption{Observed and modeled radiant positions for July $\xi$ Arietids and comet C/1964 N1 Ikeya. Rows in bold letters show radiant positions of the entries above them at 106.7° of solar longitude. The applied radiant drift was provided in the respective papers.}
        \label{tab:table4} 
        \centering 
        \begin{tabular}{l c c c c} 
        \hline\hline 
        July $\xi$ Arietids & R.A. & Dec. & V$_{g}$ & $\lambda_{0}$\\
        References & (\degr) & (\degr) & (km/s) & (\degr)\\
        \hline 
        1 & 40.1 & 10.6 & 69.4 & 119.0\\
         & \textbf{32.0} & \textbf{7.5} & \textbf{...} & \textbf{106.7}\\ 
        2 & 35.0 & 9.2 & 68.9 & 112.6\\
        3 & 33.8 & 8.7 & 68.3 & 111.1\\
        4 & 41.5 & 10.7 & 68.9 & 119.0\\
         & \textbf{29.6} & \textbf{7.0} & \textbf{...} & \textbf{106.7}\\ 
        1964 N1 Ikeya \\meteoroids, this work & 29.0 & 6.5 & 68.7 & 106.7\\
        \hline 
        \end{tabular}
        \tablebib{(1) \citet{segon2014new}; (2) \citet{kornovs2014confirmation}; (3) \citet{rudawska2015independent}; (4) \citet{jenniskens2016cams}.
        }
\end{table}

However, the radiant position for solar longitude as found from dynamical modeling fits very well with that predicted by the radiant's daily motion: assuming $\Delta$R.A. = 0.66° and $\Delta$Dec. = 0.25° from \citet{segon2014new}, the radiant position at $\lambda_{0}$ = 106.7° would be located at R.A. = 32.0°, Dec. = 7.5° or about three degrees from the modeled radiant. If we use results from \citet{jenniskens2016cams} ($\Delta$R.A. = 0.97° and $\Delta$Dec. = 0.30°), the resulting radiant position fits even better – having R.A. = 29.0° Dec. = 7.0° or about one degree from the modeled radiant. The fact that the model does not fit the observed activity may be explained by various factors, from the lack of precise data of the comet position in the past derived using the relatively small orbit arc of observations, to the possibility that this shower has some other parent body (possibly associated to C/1964 N1 Ikeya) as well. The next closest possible association was with comet 1987 B1 Nishikawa-Takamizawa-Tago where the D$_{SH}$ was 0.21, but due to high nodal distances between orbits, we consider this to be not connected to the July $\xi$ Arietids.

\begin{figure}
        \resizebox{\hsize}{!}{\includegraphics{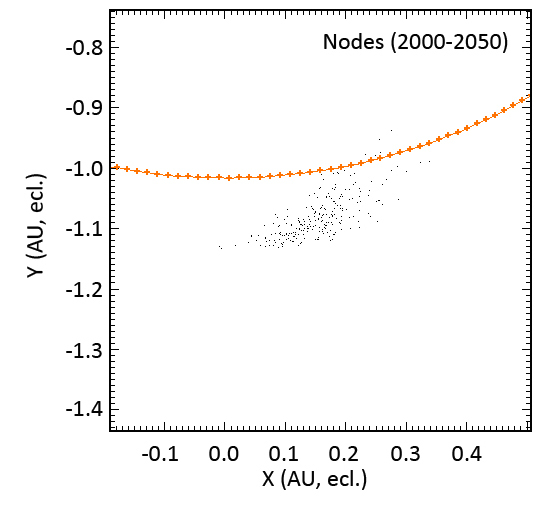}}
        \caption{Location of the particles ejected by comet C/1964 N1 Ikea over several centuries, concatenated over 50 years in the vicinity of the Earth.}
        \label{fig:figure3}
\end{figure}

\begin{figure}
        \resizebox{\hsize}{!}{\includegraphics{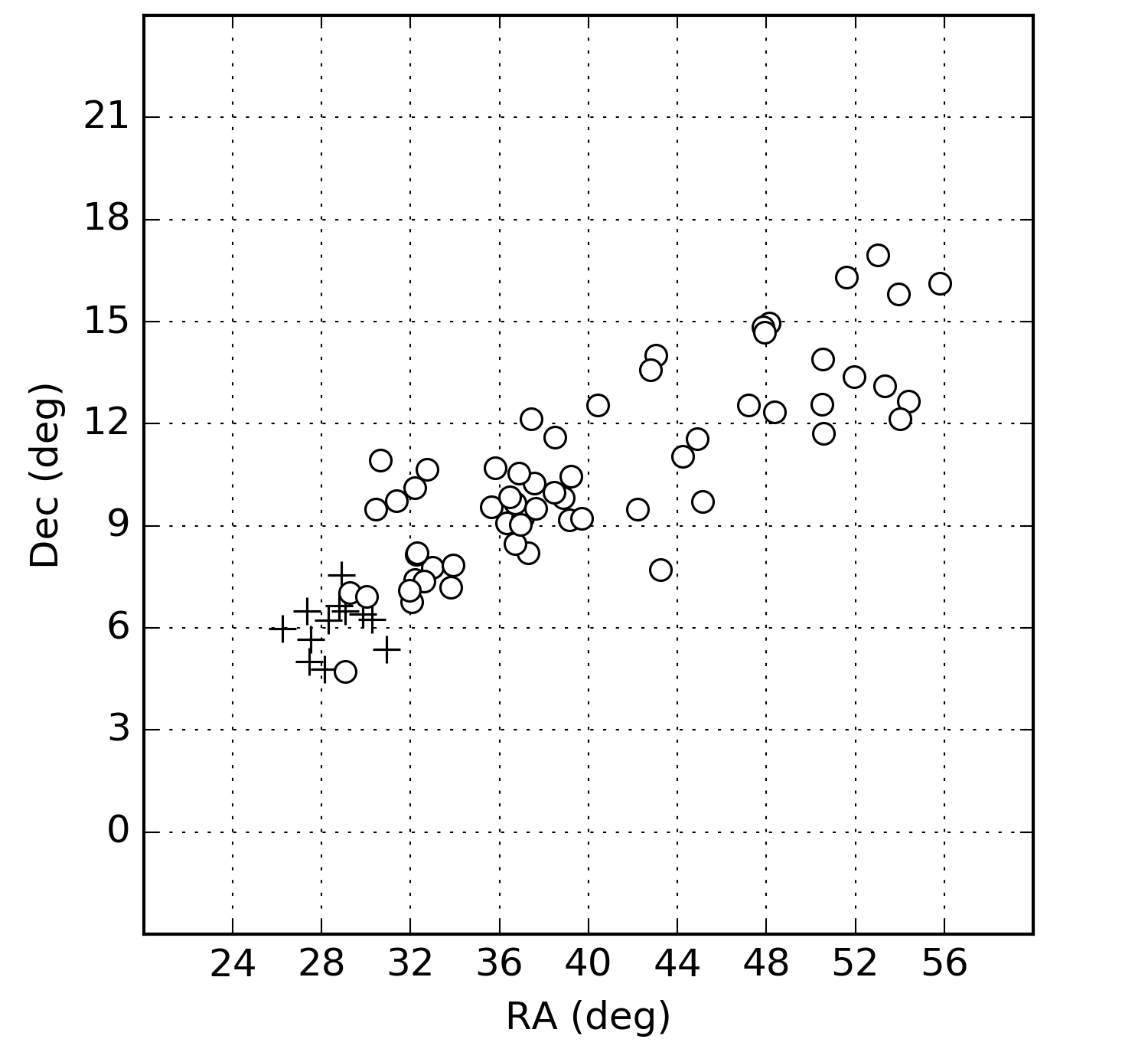}}
        \caption{Theoretical radiant of the particles released by C/1964 N1 Ikea which were closest to the Earth. The match with the July $\xi$ Arietids is not convincing in this case. The range of solar longitudes for modeled radiants is from 99.0\degr to 104.8\degr. Pluses represent the modeled radiants in the given solar longitude range, while the circles represent the observed radiants during the whole activity of the shower.}
        \label{fig:figure4}
\end{figure}

The simulation of the meteoroid stream was performed for hypothetical comet returns back to 5000 years before the present. According to the known orbit of the comet, it experienced a close encounter with Jupiter and Saturn in 1676 and 1673 AD respectively, making the orbital evolution prior to this date much more uncertain. Nevertheless, the simulation of the stream was performed in order to get a big picture view of the stream in the present day solar system as visualized in Figures \ref{fig:figure3} and \ref{fig:figure4}.

\section{IAU meteor shower \#539 ACP - $\alpha$ Cepheids and comet 255P Levy
}

The $\alpha$ Cepheids shower had been reported by \citet{segon2014new}, as a disperse grouping of 41 meteors active from mid-December to mid-January at a mean radiant position of R.A. = 318°, Dec. = 64° at $\lambda_{0}$ = 281° (January 2). The authors investigated the possibility that this new shower could be connected to the predicted enhanced meteor activity of IAU shower \#446 DPC December $\phi$ Cassiopeiids. However, the authors pointed out that \#466 DPC and \#539 ACP cannot be the same meteor shower \citep{segon2014new}. Despite the fact that a predicted meteor outburst was not detected \citep{roggemans2014letter}, there is a strong possibility that the activity from comet 255P/Levy produces a meteor shower which can be observed from the Earth as the $\alpha$ Cepheids shower. Meteor searches conducted by \citet{kornovs2014confirmation} and \citet{jenniskens2016cams} failed to detect this shower, but \citet{rudawska2015independent} found 11 meteors with a mean radiant position at R.A. = 333.5°, Dec. = +66°, V$_{g}$ = 13.4 km/s at $\lambda_{0}$ = 277.7°.

\begin{figure}[b]
        \resizebox{\hsize}{!}{\includegraphics{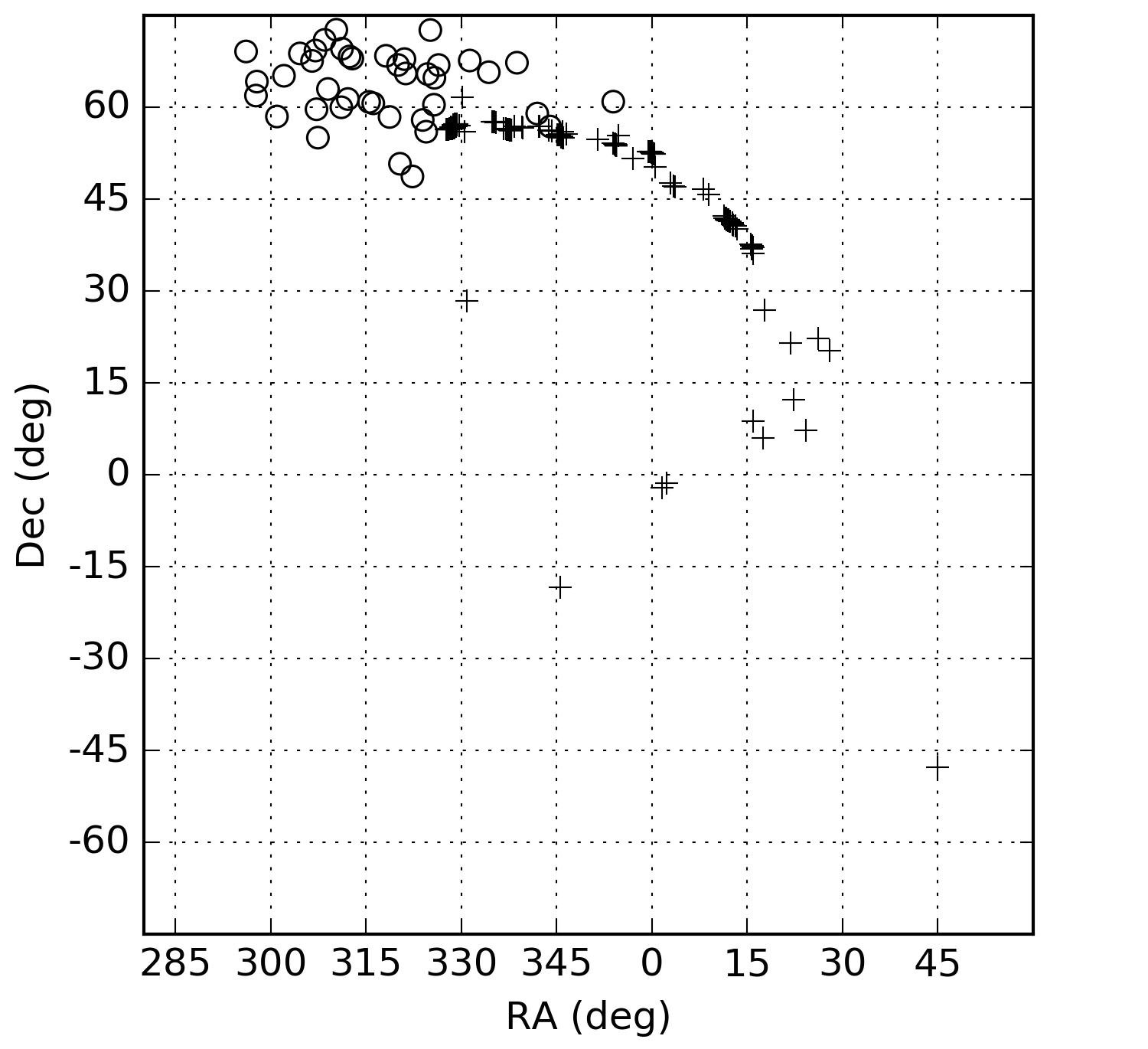}}
        \caption{Radiant positions of observed $\alpha$ Cepheids and predicted meteors from  255P/Levy. The range of solar longitudes for modeled radiants is from 250\degr to 280\degr. Pluses represent the modeled radiants in the given solar longitude range, while the circles represent the observed radiants during the whole activity of the shower.}
        \label{fig:figure5}
\end{figure}

The mean geocentric velocity for the $\alpha$ Cepheids meteors has been found to be small, of only 15.9 km/s, but ranges from 12.4 to 19.7 kilometres per second. Such a high dispersion in velocities may be explained by the fact that the D-criterion threshold for automatic search has been set to D$_{SH}$ = 0.15, which allowed a wider range of orbits to be accepted as meteor shower members. According to the dynamical modeling results, the geocentric velocity for meteoroids ejected from 255P/Levy should be of about 13 km/s, and observations show that some of the $\alpha$ Cepheids meteors indeed have such velocities at more or less the predicted radiant positions, as can be seen from Figure \ref{fig:figure5}. This leads us to the conclusion that this meteor shower has to be analyzed in greater detail, but at least some of the observations represent meteoroids coming from comet 255P/Levy. 

\begin{figure}
        \resizebox{\hsize}{!}{\includegraphics{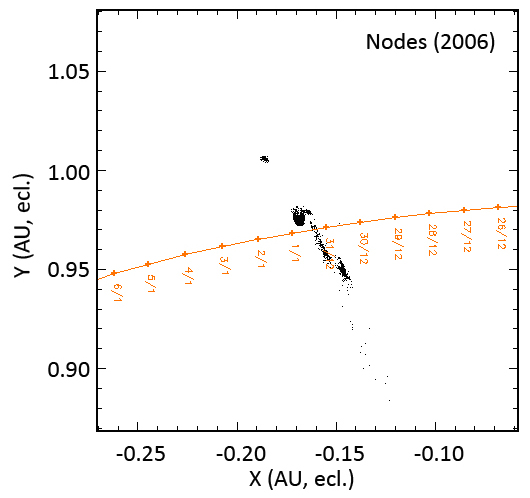}}
        \caption{Location of the particles ejected by comet 255P/Levy in the vicinity of the Earth in 2006: an outburst should have been detected.}
        \label{fig:figure6}
\end{figure}

\begin{figure}
        \resizebox{\hsize}{!}{\includegraphics{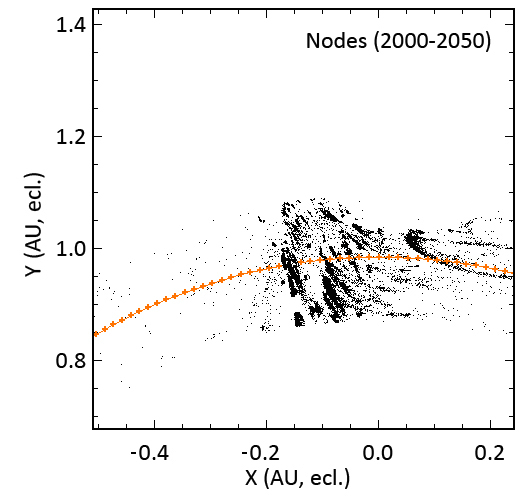}}
        \caption{Location of all the particles ejected by 255P over 50 years in order to show the location of the whole stream in the solar system. This graph does not imply several outbursts but rather provides a global indication of the stream.}
        \label{fig:figure7}
\end{figure}

The simulation of the meteoroid stream ejected by comet 255P/Levy includes trails ejected from 1801 through 2017 as visualized in Figures \ref{fig:figure6} and \ref{fig:figure7}. Several past outbursts were forecasted by the dynamical modeling but none had been observed, namely during apparitions in 2006 and 2007 (see Table \ref{tab:table5}). As a consequence, the conclusion is that the activity of the $\alpha$ Cepheids is most likely due to the global background of the stream.

\begin{table}
        \caption{Expected outburst caused by 255P/Levy. No unusual outburst was reported in 2006 and 2007. Columns: Year = the year of Earth's collision with the trail, Trail = year of particle ejection from the given trail, $\lambda_{0}$ = solar longitude in degrees, yyyy-mm-ddThh:mm:ss = date and time of the trail's closest approach, ZHR = zenithal hourly rate.}
        \label{tab:table5} 
        \centering 
        \begin{tabular}{c c c c c} 
        \hline\hline 
        Year & Trail & $\lambda_{0}$ & yyyy-mm-ddThh:mm:ss & ZHR\\
             &       & (\degr) & &\\
        \hline 
        2001 & 1963 & 279.132 & 2001-12-30T18:37:00 &  1\\
        2001 & 1975 & 279.765 & 2001-12-31T12:01:00 &  3\\
        2001 & 1980 & 279.772 & 2001-12-31T15:00:00 &  2\\
        2001 & 1985 & 279.828 & 2001-12-31T11:24:00 & 11\\
        2001 & 1991 & 279.806 & 2001-12-31T10:44:00 & 13\\
        2002 & 1963 & 278.914 & 2002-10-20T14:56:00 &  1\\
        2002 & 1980 & 279.805 & 2002-12-31T10:23:00 &  2\\
        2002 & 1985 & 279.808 & 2002-12-31T10:40:00 & 15\\
        2002 & 1991 & 279.789 & 2002-12-31T10:24:00 &  6\\
        2006 & 1963 & 279.285 & 2006-12-31T08:01:00 &  1\\
        2007 & 1963 & 279.321 & 2007-12-31T07:04:00 &  1\\
        2012 & 1980 & 279.803 & 2012-12-31T06:25:00 &  6\\
        2013 & 1980 & 279.882 & 2013-12-31T08:16:00 &  2\\
        2014 & 1969 & 264.766 & 2014-12-17T00:07:00 &  1\\
        2017 & 1930 & 342.277 & 2017-09-21T18:39:00 &  1\\
        2017 & 1941 & 279.510 & 2017-12-30T03:41:00 &  1\\
        2018 & 1969 & 278.254 & 2018-12-29T07:29:00 &  1\\
        2033 & 1975 & 275.526 & 2033-12-27T10:12:00 &  1\\
        2033 & 1980 & 275.488 & 2033-12-27T10:06:00 &  1\\
        2033 & 1985 & 275.452 & 2033-12-27T09:55:00 &  1\\
        2033 & 1991 & 275.406 & 2033-12-27T09:54:00 &  1\\
        2033 & 1996 & 275.346 & 2033-12-27T08:58:00 &  1\\
        2034 & 1975 & 262.477 & 2034-12-13T22:22:00 &  1\\
        2034 & 1980 & 261.456 & 2034-06-06T03:40:00 &  1\\
        2034 & 1985 & 261.092 & 2034-04-05T17:02:00 &  1\\
        2034 & 1991 & 260.269 & 2034-03-09T11:52:00 &  1\\
        2035 & 1914 & 276.553 & 2035-01-09T07:59:00 &  1\\
        2035 & 1952 & 271.463 & 2035-12-20T03:11:00 &  1\\
        2039 & 1980 & 272.974 & 2039-12-25T01:51:00 &  1\\
        2039 & 1991 & 272.131 & 2039-12-25T01:05:00 &  1\\
        \hline 
        \end{tabular}
\end{table}

There are several other parent bodies possibly connected to the $\alpha$ Cepheids stream: 2007 YU56 (D$_{SH}$ = 0.20),  2005 YT8 (D$_{SH}$ = 0.19),  1999 AF4 (D$_{SH}$ = 0.19),  2011 AL52 (D$_{SH}$ = 0.19),  2013 XN24 (D$_{SH}$ = 0.12),  2008 BC (D$_{SH}$ = 0.17),  and  2002 BM (D$_{SH}$ = 0.16). The analysis for those bodies will be done in a future analysis.

\section{IAU meteor shower \#541 SSD - 66 Draconids and asteroid 2001 XQ
}
Meteor shower 66 Draconids had been reported by \citet{segon2014new}, as a grouping of 43 meteors having mean radiant at R.A. = 302°, Dec. = +62°, V$_{g}$ = 18.2 km/s. This shower has been found to be active from solar longitude 242° to 270° (November 23 to December 21), having a peak activity period around 255° (December 7). Searches by \citet{jenniskens2016cams} and \citet{kornovs2014confirmation} failed to detect this shower. But again, \citet{rudawska2015independent} found this shower to consist of 39 meteors from the EDMOND meteor orbits database, at R.A. = 296°, Dec. = 64°, V$_{g}$ = 19.3 km/s for solar longitude $\lambda_{0}$ = 247°.

A search for a possible parent body of this shower resulted in asteroid 2001 XQ, which having a D$_{SH}$ = 0.10 represented the most probable choice. The summary of mean orbital parameters from the above mentioned searches compared with 2001 XQ are shown in Table \ref{tab:table6}. 

\begin{table*}[t]
        \caption{Orbital parameters for 66 Draconids and 2001XQ with respective D$_{SH}$ values. Orbital elements (mean values for shower data): q = perihelion distance, e = eccentricity, i = inclination, Node = Node, $\omega$ = argument of perihelion, D$_{SH}$ = Southworth and Hawking D-criterion with respect to 2001XQ.}
        \label{tab:table6} 
        \centering 
        \begin{tabular}{l c c c c c c} 
        \hline\hline 
        66 Draconids & q & e & i & Node & $\omega$ & D$_{SH}$\\
        References & (AU) & & (\degr) & (\degr) & (\degr) & \\
        \hline 
        1 & 0.981 & 0.657 & 27.2 & 255.2 & 184.4 & 0.10\\
        2 & 0.980 & 0.667 & 29.0 & 247.2 & 185.2 & 0.13\\
        2001 XQ & 1.035 & 0.716 & 29.0 & 251.4 & 190.1 & 0\\
        \hline 
        \end{tabular}
        \tablebib{(1) \citet{segon2014new}; (2) \citet{rudawska2015independent}.
        }
\end{table*}

Asteroid 2001 XQ has Tisserand parameter T$_{j}$ = 2.45, which is a value common for Jupiter family comets and this makes us suspect it may not be an asteroid per se, but rather a dormant comet. To the collected author's knowledge, no cometary activity has been observed for this body. Nor was there any significant difference in the full-width half-max spread between stars and the asteroid on the imagery provided courtesy of Leonard Kornoš (personal communication) from Modra Observatory. They had observed this asteroid (at that time named 2008 VV4) on its second return to perihelion, during which it reached \nth{18} magnitude.

Numerical modeling of the hypothetical meteor shower whose particles originate from asteroid 2001 XQ was performed for perihelion passages from 800 AD up to 2100 AD. The modeling showed multiple direct hits into the Earth for many years, even outside the period covered by the observations. The summary of observed and modeled radiant positions is given in Table \ref{tab:table7}. 

\begin{table}
        \caption{Observed 66 Draconid and modeled 2001 XQ meteors' mean radiant positions (prefix C\_ stands for calculated (modeled), while prefix O\_ stands for observed). The number in the parenthesis indicates the number of observed 66 Draconid meteors in the given year. $\theta$ is the angular distance between the modeled and the observed mean radiant positions.}
        \label{tab:table7} 
        \centering 
        \begin{tabular}{l c c c c c} 
        \hline\hline 
        Year & $\lambda_{0}$ & R.A. & Dec. & V$_{g}$ & $\theta$\\
         & (\degr) & (\degr) & (\degr) & (km/s) & (\degr)\\
        \hline 
        C\_2007 & 250.3 & 308.2 & 65.3 & 19.3 & ...\\
        O\_2007 (5) & 257.5 & 300.1 & 63.2 & 18.2 & 4.1\\
        C\_2008 & 248.2 & 326.8 & 56.9 & 16.1 & ...\\
        O\_2008 (8) & 254.0 & 300.5 & 62.6 & 18.0 & 14.3\\
        C\_2009 & 251.1 & 309.6 & 64.0 & 18.8 & ...\\
        O\_2009 (5) & 253.6 & 310.4 & 61.0 & 17.0 & 3.0\\
        C\_2010 & 251.2 & 304.0 & 63.1 & 19.1 & ...\\
        O\_2010 (17) & 253.7 & 300.4 & 63.4 & 18.9 & 1.6\\
        \hline 
        \end{tabular}
\end{table}

\begin{figure}
        \resizebox{\hsize}{!}{\includegraphics{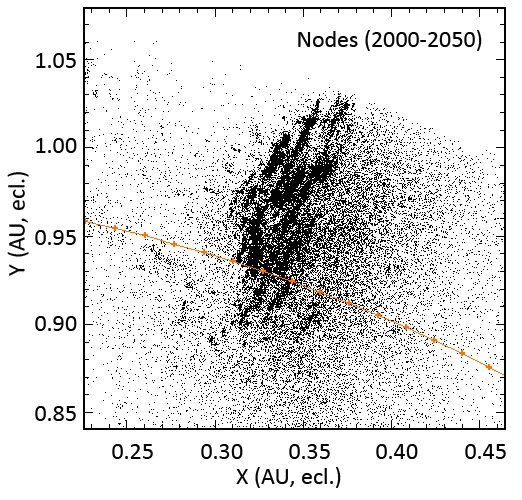}}
        \caption{Location of the nodes of the particles released by 2001 XQ over several centuries, concatenated over 50 years. The Earth crosses the stream.}
        \label{fig:figure8}
\end{figure}

\begin{figure}
        \resizebox{\hsize}{!}{\includegraphics{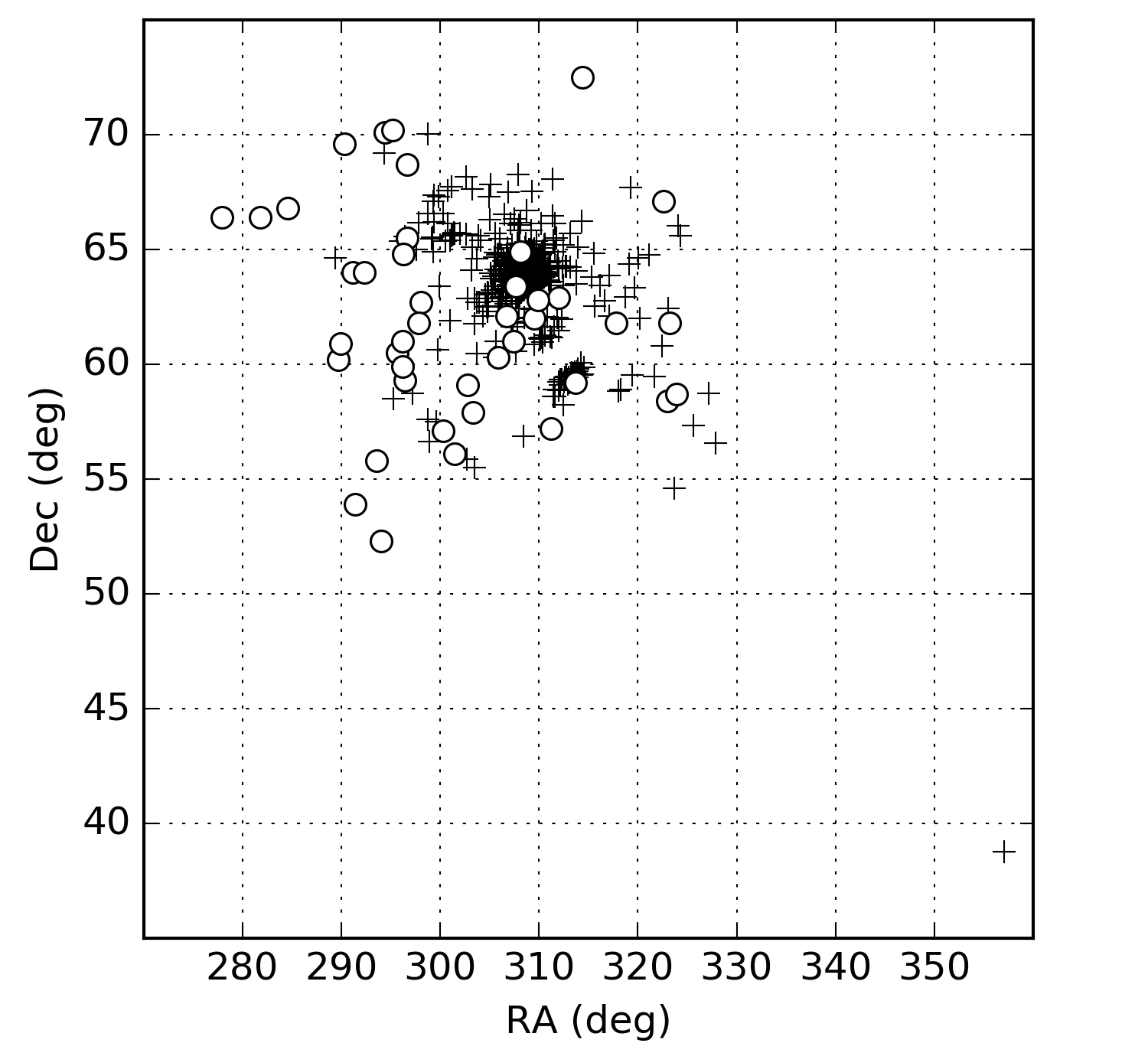}}
        \caption{Theoretical radiants of the particles released by 2001 XQ which were closest to the Earth. The range of solar longitudes for modeled radiants is from 231.1\degr to 262.8\degr. Pluses represent the modeled radiants in the given solar longitude range, while the circles represent the observed radiants during the whole activity of the shower.}
        \label{fig:figure9}
\end{figure}

Despite the fact that the difference in the mean radiant positions may seem significant, radiant plots of individual meteors show that some of the meteors predicted to hit the Earth at the observation epoch were observed at positions almost exactly as predicted. It is thus considered that the results of the simulations statistically represent the stream correctly, but individual trails cannot be identified as responsible for any specific outburst, as visualized in Figures \ref{fig:figure8} and \ref{fig:figure9}. The activity of this shower is therefore expected to be quite regular from year to year.

Two other candidate parent bodies were initially considered, 2004 YY23 and 2015 WB13, in which both had a D$_{SH}$ of 0.26. This was deemed too distant to be associated with the 66 Draconids stream.

\section{IAU meteor shower \#751 KCE - $\kappa$ Cepheids and asteroid 2009 SG18
}

The meteor shower $\kappa$ Cepheids had been reported by \citet{segon2015four}, as a grouping of 17 meteors with very similar orbits, having average D$_{SH}$ of only 0.06. The activity period was found to from September 11 to September 23, covering solar longitudes from 168° to 180°. The radiant position was R.A. = 318°, Dec. = 78° with V$_{g}$ = 33.7 km/s, at a mean solar longitude of 174.4°. Since the new shower discovery has been reported only recently, the search by \citet{kornovs2014confirmation} could be considered totally blind having not found its existence, while the search by \citet{jenniskens2016cams} did not detect it as well in the CAMS database. Once again, the search by \citet{rudawska2015independent} found the shower, but in much higher numbers than it has been found in the SonotaCo and CMN orbit databases. In total 88 meteors have been extracted as $\kappa$ Cepheids members in the EDMOND database. A summary of the mean orbital parameters from the above mentioned searches compared with 2009 SG18 are shown in Table \ref{tab:table8}.

\begin{table*}[t]
        \caption{Orbital parameters for $\kappa$ Cepheids and asteroid 2009 SG18 with corresponding D$_{SH}$ values. Orbital elements (mean values for shower data): q = perihelion distance, e = eccentricity, i = inclination, Node = Node, $\omega$ = argument of perihelion, D$_{SH}$ = Southworth and Hawking D-criterion with respect to 2009 SG18.}
        \label{tab:table8} 
        \centering 
        \begin{tabular}{l c c c c c c} 
        \hline\hline 
        $\kappa$ Cepheids & q & e & i & Node & $\omega$ & D$_{SH}$\\
        References & (AU) & & (\degr) & (\degr) & (\degr) & \\
        \hline 
        1 & 0.983 & 0.664 & 57.7 & 174.4 & 198.4 & 0.10\\
        2 & 0.987 & 0.647 & 55.9 & 177.2 & 190.4 & 0.17\\
        2009 SG18 & 0.993 & 0.672 & 58.4 & 177.6 & 204.1 & 0\\
        \hline 
        \end{tabular}
        \tablebib{(1) \citet{segon2014new}; (2) \citet{rudawska2015independent}.
        }
\end{table*}

What can be seen at a glance is that the mean orbital parameters for both searches are very consistent (D$_{SH}$ = 0.06), while the difference between the mean shower orbits and the asteroid's orbit differs mainly in the argument of perihelion and perihelion distance. Asteroid 2009 SG18 has a Tisserand parameter for Jupiter of T$_{j}$ = 2.31, meaning that it could be a dormant comet.

\begin{figure}
        \resizebox{\hsize}{!}{\includegraphics{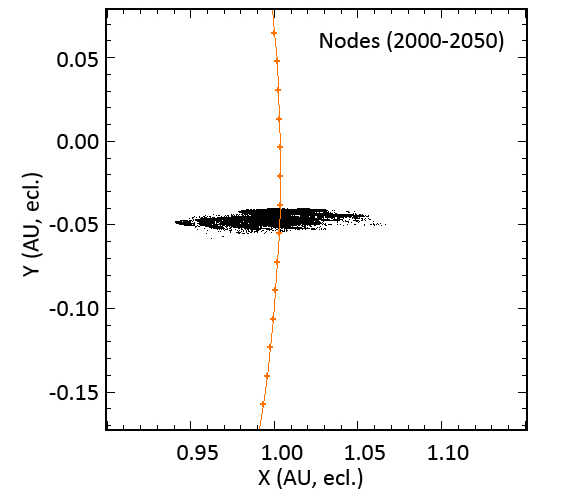}}
        \caption{Location of the nodes of the particles released by 2009 SG18 over several centuries, concatenated over 50 years. The Earth crosses the stream.}
        \label{fig:figure10}
\end{figure}

\begin{figure}
        \resizebox{\hsize}{!}{\includegraphics{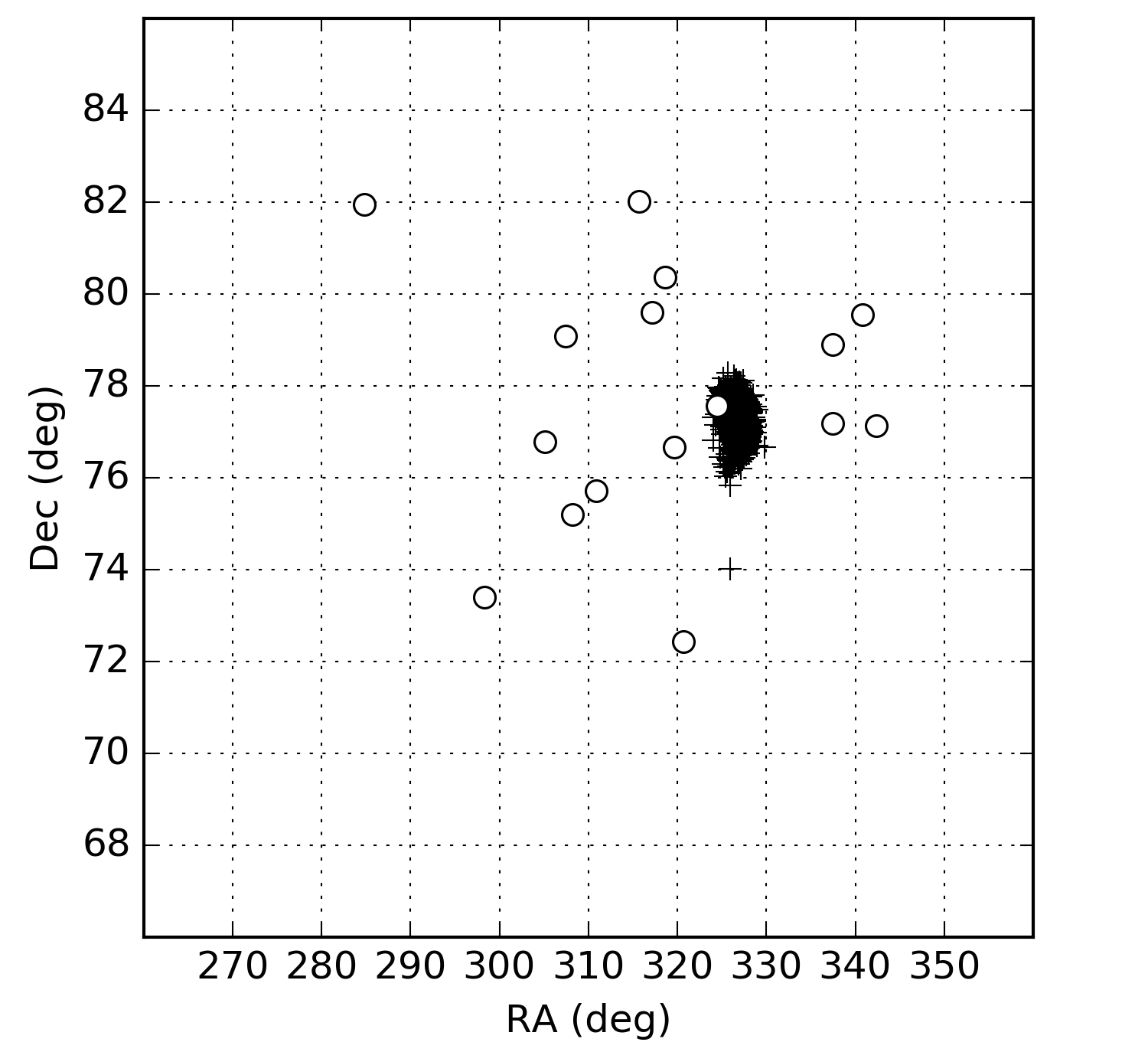}}
        \caption{Theoretical radiant of the particles released by 2009 SG18 which were closest to the Earth. Several features are visible due to the difference trails, but care must be taken when interpreting these data. The range of solar longitudes for modeled radiants is from 177.0\degr to 177.7\degr. Pluses represent the modeled radiants in the given solar longitude range, while the circles represent the observed radiants during the whole activity of the shower.}
        \label{fig:figure11}
\end{figure}

Numerical modeling of the hypothetical meteor shower originating from asteroid 2009 SG18 for perihelion passages from 1804 AD up to 2020 AD yielded multiple direct hits into the Earth for more years than the period covered by the observations, as seen in Figures \ref{fig:figure10} and \ref{fig:figure11}. The very remarkable coincidence found between the predicted and observed meteors for years 2007 and 2010 is summarized in Table \ref{tab:table9}.

\begin{table}
        \caption{Observed $\kappa$ Cepheids and modeled 2009 SG18 meteors' mean radiant positions (prefix C\_ stands for calculated or modeled, while prefix O\_ stands for observed). The number in the parenthesis indicates the number of observed meteors in the given year. $\theta$ is the angular distance between the modeled and the observed mean radiant positions.}
        \label{tab:table9} 
        \centering 
        \begin{tabular}{l c c c c c} 
        \hline\hline 
        Year & $\lambda_{0}$ & R.A. & Dec. & V$_{g}$ & $\theta$\\
         & (\degr) & (\degr) & (\degr) & (km/s) & (\degr)\\
        \hline 
        C\_2007 & 177.4 & 327.5 & 77.0 & 34.0 & ...\\
        O\_2007 (3) & 177.1 & 328.3 & 77.9 & 35.3 & 0.9\\
        C\_2010 & 177.7 & 327.7 & 77.7 & 34.3 & ...\\
        O\_2010 (2) & 177.7 & 326.5 & 80.5 & 34.7 & 2.8\\
        \hline 
        \end{tabular}
\end{table}

Based on an initial analysis given in this paper, a prediction of possible enhanced activity on September 21, 2015 was made by \citet{segon2015croatian}. At the moment, there are no video meteor data that confirm the prediction of the enhanced activity, but a paper on visual observations of the $\kappa$ Cepheids by a highly reputable visual observer confirmed some level of increased activity (\citet{rendtel2015minor}).

The encounters shown in Table \ref{tab:table10} between the trails ejected by 2009 SG18 and the Earth were found theoretically through the dynamical modeling. Caution should be emphasized when interpreting the results since the confirmation of any historical outbursts still need to be performed before trusting such predictions.

\begin{table*}[t]
        \caption{Prediction of possible outbursts caused by 2009 SG18. Columns: Year = the year of Earth's collision with the trail, Trail = year of particle ejection from the given trail, rE-rD = the distance between the Earth and the center of the trail, $\lambda_{0}$ = solar longitude in degrees, yyyy-mm-ddThh:mm:ss = date and time of the trail's closest approach, ZHR = zenithal hourly rate}
        \label{tab:table10} 
        \centering 
        \begin{tabular}{c c c c c c} 
        \hline\hline 
        Year & Trail & rE-rD & $\lambda_{0}$ & yyyy-mm-ddThh:mm:ss & ZHR\\
        & & (AU) & (\degr) & &\\
        \hline 
        2005 & 1967 &  0.00066 & 177.554 & 2005-09-20T12:08:00 & 11\\
        2006 & 1804 &  0.00875 & 177.383 & 2006-09-20T11:31:00 & 13\\
        2010 & 1952 & -0.00010 & 177.673 & 2010-09-20T21:38:00 & 12\\
        2015 & 1925 & -0.00143 & 177.630 & 2015-09-21T03:29:00 & 10\\
        2020 & 1862 & -0.00064 & 177.479 & 2020-09-20T06:35:00 & 11\\
        2021 & 1962 &  0.00152 & 177.601 & 2021-09-20T15:39:00 & 11\\
        2031 & 2004 & -0.00126 & 177.267 & 2031-09-20T21:15:00 & 12\\
        2031 & 2009 & -0.00147 & 177.222 & 2031-09-20T19:55:00 & 13\\
        2033 & 1946 &  0.00056 & 177.498 & 2033-09-20T14:57:00 & 10\\
        2036 & 1978 & -0.00042 & 177.308 & 2036-09-20T04:44:00 & 20\\
        2036 & 2015 & -0.00075 & 177.220 & 2036-09-20T02:33:00 & 20\\
        2036 & 2025 &  0.00109 & 177.254 & 2036-09-20T03:19:00 & 13\\
        2037 & 1857 & -0.00031 & 177.060 & 2037-09-20T04:37:00 & 13\\
        2037 & 1946 &  0.00021 & 177.273 & 2037-09-20T09:56:00 & 10\\
        2038 & 1841 & -0.00050 & 177.350 & 2038-09-20T18:02:00 & 10\\
        2038 & 1925 &  0.00174 & 177.416 & 2038-09-20T19:39:00 & 11\\
        2039 & 1815 & -0.00018 & 177.303 & 2039-09-20T23:01:00 & 10\\
        \hline 
        \end{tabular}
\end{table*}

The next closest possible association was 2002 CE26 with D$_{SH}$ of 0.35, which was deemed too distant to be connected to the $\kappa$ Cepheids stream.

\section{IAU meteor shower \#753 NED  - November Draconids and asteroid 2009 WN25
}

The November Draconids had been previously reported by \citet{segon2015four}, and consist of 12 meteors on very similar orbits having a maximal distance from the mean orbit of D$_{SH}$ = 0.08, and on average only D$_{SH}$ = 0.06. The activity period was found to be between November 8 and 20, with peak activity at solar longitude of 232.8°. The radiant position at peak activity was found to be at R.A. = 194°, Dec. = +69°, and V$_{g}$ = 42.0 km/s. There are no results from other searches since the shower has been reported only recently. Other meteor showers were reported on coordinates similar to \#753 NED, namely \#387 OKD October $\kappa$ Draconids and \#392 NID November i Draconids \citep{brown2010meteoroid}. The difference in D$_{SH}$ for \#387 OKD is found to be far too excessive (0.35) to be considered to be the same shower stream. \#392 NID may be closely related to \#753, since the D$_{SH}$ of 0.14 derived from radar observations show significant similarity; however, mean orbits derived from optical observations by \citet{jenniskens2016cams} differ by D$_{SH}$ of 0.24 which we consider too far to be the same shower.

\begin{figure}
        \resizebox{\hsize}{!}{\includegraphics{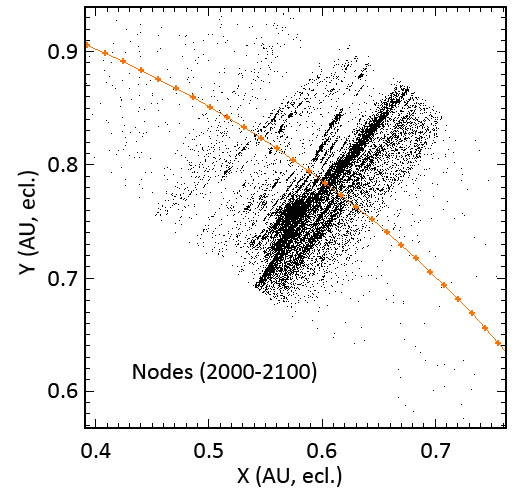}}
        \caption{Location of the nodes of the particles released by 2009 WN25 over several centuries, concatenated over 100 years. The Earth crosses the stream.}
        \label{fig:figure12}
\end{figure}

\begin{figure}
        \resizebox{\hsize}{!}{\includegraphics{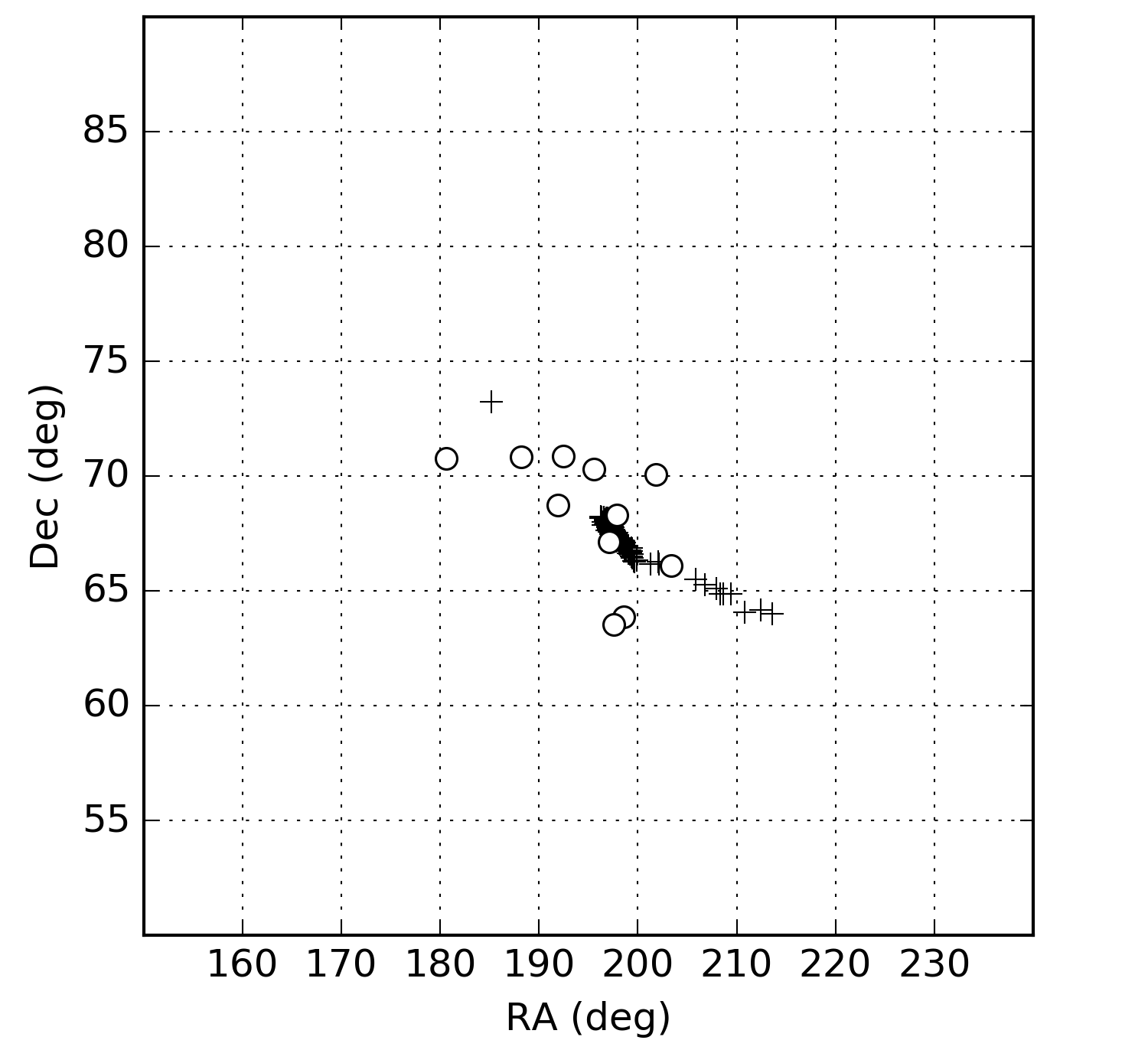}}
        \caption{Theoretical radiant of the particles released by 2009 WN25 which were closest to the Earth. The range of solar longitudes for modeled radiants is from 230.3\degr to 234.6\degr. Pluses represent the modeled radiants in the given solar longitude range, while the circles represent the observed radiants during the whole activity of the shower.}
        \label{fig:figure13}
\end{figure}

The possibility that asteroid 2009 WN25 is the parent body of this possible meteor shower has been investigated by numerical modeling of the hypothetical meteoroids ejected for the period from 3000 BC up to 1500 AD and visualized in Figures \ref{fig:figure12} and \ref{fig:figure13}. The asteroid 2009 WN25 has a Tisserand parameter for Jupiter of T$_{j}$ = 1.96. Despite the fact that direct encounters with modeled meteoroids were not found for all years in which the meteors were observed, and that the number of hits is relatively small compared to other modeled showers, the averaged predicted positions fit the observations very well (see Table \ref{tab:table11}). This shows that the theoretical results have a statistically meaningful value and validates the approach of simulating the stream over a long period of time and concatenating the results to provide an overall view of the shower.

\begin{table}
        \caption{Averaged observed and modeled radiant positions for \#753 NED and 2009 WN25. $\theta$ is the angular distance between the modeled and the observed mean radiant positions.}
        \label{tab:table11} 
        \centering 
        \begin{tabular}{l c c c c c} 
        \hline\hline 
        November Draconids & $\lambda_{0}$ & R.A. & Dec. & V$_{g}$ & $\theta$\\
         & (\degr) & (\degr) & (\degr) & (km/s) & (\degr)\\
        \hline 
        Predicted & 232.8 & 194.2 & 68.6 & 42.0 & ...\\
        Observed & 232.4 & 196.5 & 67.6 & 41.8 & 1.3\\
        \hline 
        \end{tabular}
\end{table}

Moreover, it appears that the predicted 2014 activity sits exactly at the same equatorial location (R.A. = 199°, Dec. = +67°) seen on Canadian Meteor Orbit Radar (CMOR) plots.\footnote{\label{cmor_plots}\url{http://fireballs.ndc.nasa.gov/} - "radar".} The shower has been noted as NID, but its position fits more closely to the NED. Since orbital data from the CMOR database are not available online, the authors were not able to confirm the hypothesis that the radar is seeing the same meteoroid orbits as the model produces. However, the authors received a confirmation from Dr. Peter Brown at the University of Western Ontario (private correspondence) that this stream has shown activity each year in the CMOR data, and likely belongs to the QUA-NID complex. A recently published paper \citep{micheli2016evidence} suggests that asteroid 2009 WN25 may be a parent body of the NID shower as well, so additional analysis with more observations will be needed to reveal the true nature of this shower complex. The next closest possible association was 2012 VF6 with D$_{SH}$ of 0.49, which was deemed too distant to be connected to the November Draconids stream.

\section{IAU meteor shower \#754 POD - $\psi$ Draconids and asteroid 2008 GV
}
The possible new meteor shower $\psi$ Draconids was reported by \citet{segon2015four}, consisting of 31 tight meteoroid orbits, having maximal distance from a mean orbit of D$_{SH}$ = 0.08, and on average only D$_{SH}$ = 0.06. The $\psi$ Draconids were found to be active from March 19 to April 12, with average activity around solar longitude of 12° with radiant at R.A. = 262°, Dec. = +73°, and V$_{g}$ = 19.8 km/s. No confirmation from other shower searches exists at the moment, since the shower has been reported upon only recently.

If this shower's existence could be confirmed, the most probable parent body known at this time would be asteroid 2008 GV. This asteroid was found to have a very similar orbit to the average orbit of the $\psi$ Draconids, D$_{SH}$ being of 0.08. Since the asteroid has a Tisserand parameter of T$_{j}$ = 2.90, it may be a dormant comet as well. Dynamical modeling has been done for hypothetical meteoroids ejected for perihelion passages from 3000 BC to 2100 AD, resulting in direct hits with the Earth for almost every year from 2000 onwards.

\begin{figure}
        \resizebox{\hsize}{!}{\includegraphics{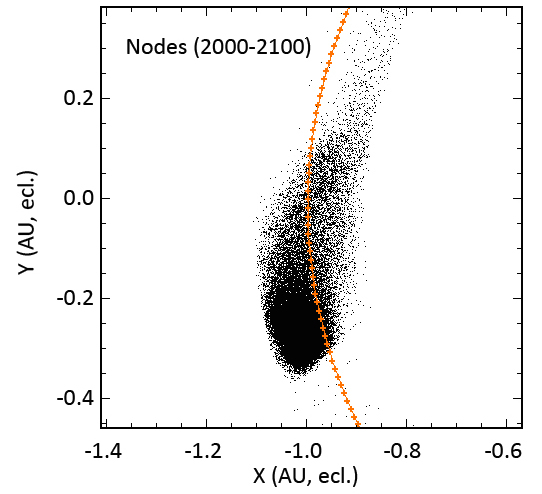}}
        \caption{Location of the nodes of the particles released by 2008 GV over several centuries, concatenated over 100 years. The Earth crosses the stream.}
        \label{fig:figure14}
\end{figure}

\begin{figure}
        \resizebox{\hsize}{!}{\includegraphics{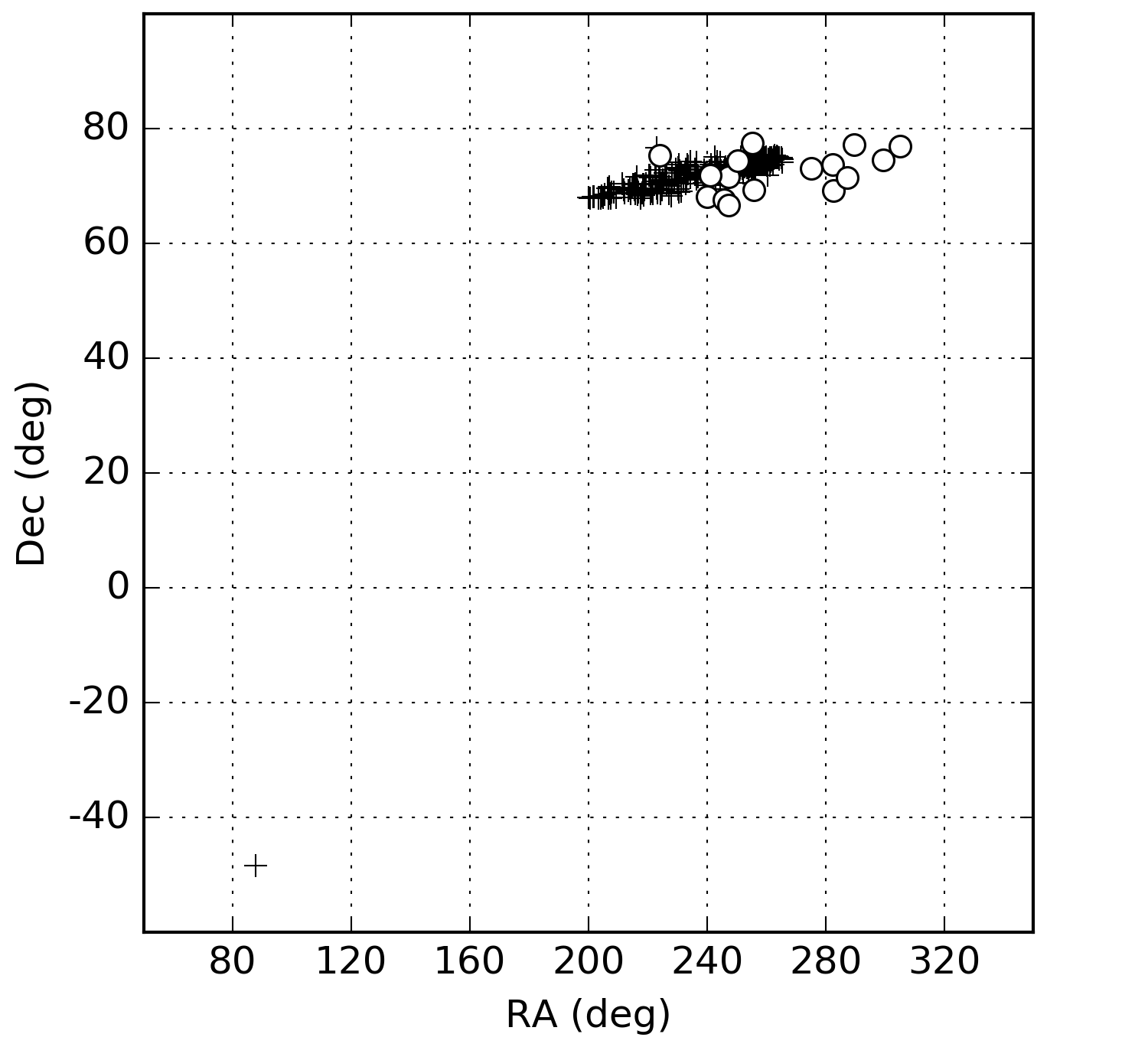}}
        \caption{Theoretical radiant of the particles released by 2008 GV which were closest to the Earth. The range of solar longitudes for modeled radiants is from 355.1\degr to 17.7\degr. Pluses represent the modeled radiants in the given solar longitude range, while the circles represent the observed radiants during the whole activity of the shower.}
        \label{fig:figure15}
\end{figure}

For the period covered by observations used in the CMN search, direct hits were found for years 2008, 2009, and 2010. The summary of the average radiant positions from the observations and from the  predictions are given in Table \ref{tab:table12}. The plots of  modeled and observed radiant positions are shown in Figure \ref{fig:figure15}, while locations of nodes of the modeled particles released by 2008 GV are shown in Figure \ref{fig:figure14}.

\begin{table}
        \caption{Observed $\psi$ Draconids and modeled 2008 GV meteors' mean radiant positions (prefix C\_ stands for calculated (modeled), while prefix O\_ stands for observed). The number in the parenthesis indicates the number of observed meteors in the given year. $\theta$ is the angular distance between the modeled and the observed mean radiant positions.}
        \label{tab:table12} 
        \centering 
        \begin{tabular}{l c c c c c} 
        \hline\hline 
        Year & $\lambda_{0}$ & R.A. & Dec. & V$_{g}$ & $\theta$\\
         & (\degr) & (\degr) & (\degr) & (km/s) & (\degr)\\
        \hline 
        C\_2008 & 15.9 & 264.6 & 75.2 & 19.4 & ...\\
        O\_2008 (2) & 14.2 & 268.9 & 73.3 & 20.7 & 2.2\\
        C\_2009 & 13.9 & 254.0 & 74.3 & 19.3 & ...\\
        O\_2009 (11) & 9.5 & 257.4 & 72.0 & 19.9 & 2.5\\
        C\_2010 & 12.8 & 244.7 & 73.4 & 19.1 & ...\\
        O\_2010 (6) & 15.1 & 261.1 & 73.0 & 19.8 & 4.7\\
        \hline 
        \end{tabular}
\end{table}

As can be seen from Table \ref{tab:table12}, the mean observations fit very well to the positions predicted by dynamical modeling, and for two cases there were single meteors very close to predicted positions. On the other hand, the predictions for year 2015 show that a few meteoroids should hit the Earth around solar longitude 14.5° at R.A. = 260°, Dec. = +75°, but no significant activity has been detected in CMN observations. However, small groups of meteors can be seen on CMOR plots for that solar longitude at a position slightly lower in declination, but this should be verified using radar orbital measurements once available. According to Dr. Peter Brown at the University of Western Ontario (private correspondence), there is no significant activity from this shower in the CMOR orbital data.

One other potential parent body may be connected to the $\psi$ Draconids stream: 2015 FA118. The analysis for that potential parent alternative will be done in a future analysis.

\section{IAU meteor shower \#755 MID - May $\iota$ Draconids and asteroid 2006 GY2
}
The possible new meteor shower May $\iota$ Draconids was reported by \citet{segon2015four}, consisting of 19 tight meteoroid orbits, having maximal distance from their mean orbit of D$_{SH}$ = 0.08, and on average only D$_{SH}$ = 0.06. The May $\iota$ Draconids were found to be active from May 7 to June 6, with peak activity around solar longitude of 60° at R.A. = 231°, Dec. = +53°, and V$_{g}$ = 16.7 km/s. No confirmation from other searches exists at the moment, since the shower has been reported in the literature only recently. Greaves (from the meteorobs mailing-list archives\footnote{\label{meteorobs}\url{http://lists.meteorobs.org/pipermail/meteorobs/2015-December/018122.html}.}) stated that this shower should be the same as \#273 PBO $\phi$ Bootids. However, if we look at the details of this shower as presented in \citet{jenniskens2006meteor}, we find that the solar longitude stated in the IAU Meteor Data Center does not correspond to the mean ascension node for three meteors chosen to represent the $\phi$ Bootid shower. If a weighted orbit average of all references is calculated, the resulting D$_{SH}$ from MID is 0.18 which we consider a large enough value to be a separate shower (if the MID exists at all). Three \#273 PBO orbits from the IAU MDC do indeed match \#755 MID, suggesting that these two possible showers may somehow be connected.

\begin{figure}
        \resizebox{\hsize}{!}{\includegraphics{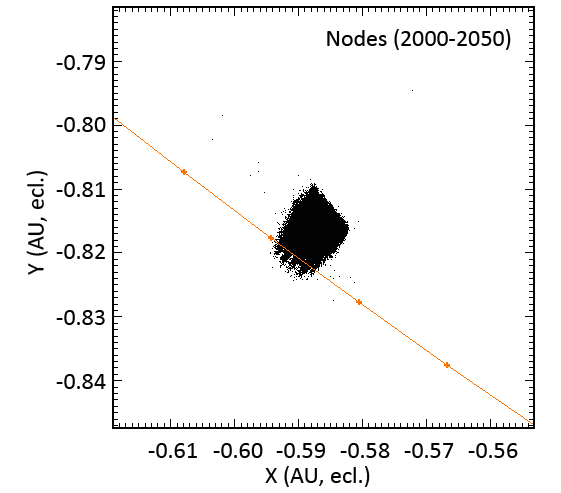}}
        \caption{Location of the nodes of the particles released by 2006 GY2 over several centuries, concatenated over 50 years. The Earth crosses the stream.}
        \label{fig:figure16}
\end{figure}

\begin{figure}
        \resizebox{\hsize}{!}{\includegraphics{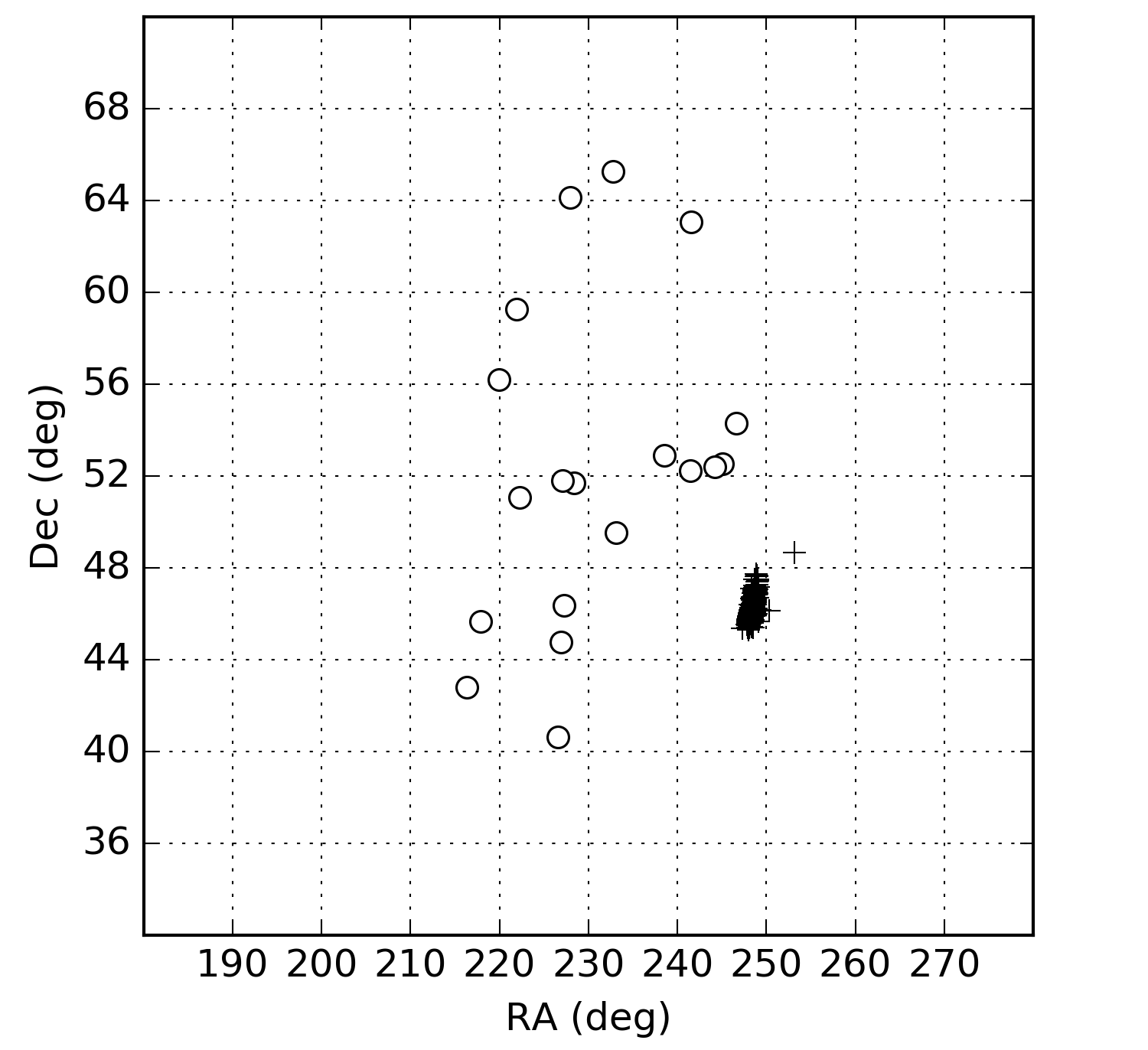}}
        \caption{Theoretical radiant of the particles released by 2006 GY2 which were closest to the Earth. The range of solar longitudes for modeled radiants is from 54.1\degr to 54.5\degr. Pluses represent the modeled radiants in the given solar longitude range, while the circles represent the observed radiants during the whole activity of the shower.}
        \label{fig:figure17}
\end{figure}

Asteroid 2006 GY2 was investigated as a probable parent body using dynamical modeling as in previous cases. The asteroid 2006 GY2 has a Tisserand parameter for Jupiter of T$_{j}$ = 3.70. From all the cases we discussed in this paper, this one shows the poorest match between the observed and predicted radiant positions. The theoretical stream was modeled with trails ejected from 1800 AD through 2100 AD. According to the dynamical modeling analysis, this parent body should produce meteors for all years covered by the observations and at more or less the same position, R.A. = 248.5°, Dec. = +46.2°, and at same solar longitude of 54.4° with V$_{g}$ = 19.3 km/s, as visualized in Figures \ref{fig:figure16} and \ref{fig:figure17}. 

However, six meteors belonging to the possible \#755 MID shower found in the solar longitude range from 52.3 to 53.8 (the next meteor found was at 58.6°) show a mean radiant position at R.A. = 225.8°, Dec. = +46.4°, with mean V$_{g}$ of 16.4 km/s. Given the angular distance of 15.6\degr from the observed radiant, the difference in geocentric velocity (3 km/s) compared to the modeled meteor radiant parameters, and the fact that there were no single model meteors observed at that position nor nearby, we cannot conclude that this asteroid may be the parent body of the possible meteor shower May $\iota$ Draconids.

Another potential parent body was 2002 KK3, having a D$_{SH}$ = 0.18. However, the dynamical modeling for 2002 KK3 showed no crossings with Earth's orbit. There were also three more distant bodies at D$_{SH}$ = 0.20:  2010 JH3, 2013 JY2, and 2014 WC7. The analysis for those bodies will be done in a future analysis.

\section{IAU meteor shower \#531 GAQ - $\gamma$ Aquilids and comet C/1853G1 (Schweizer), and other investigated bodies
}
The possible new meteor shower $\gamma$ Aquilids was reported by \citet{segon2014new}, and found in other stream search papers (\citet{kornovs2014confirmation}, \citet{rudawska2015independent} and \citet{jenniskens2016cams}). Meteoroids from the suggested parent body comet C/1853G1 (Schweizer) were modeled for perihelion passages ranging from 3000 BC up to the present, and were evaluated. Despite their being very similar orbits between \#531 GAQ and the comet C/1853G1 (Schweizer), no direct hits to the Earth were found.
Besides C/1853G1 (Schweizer), negative results were found for dynamical analyses done on asteroids 2011 YX62 (as a possible parent body of \#604 ACZ $\zeta$1 Cancrids) and 2002 KK3, 2008 UZ94, and 2009 CR2 (no shower association reported).

\section{Discussion}
\paragraph{}
The new meteoroid stream discoveries described in this work have been reported on previously, and were searches based on well-defined conditions and constraints that individual meteoroid orbits must meet for association. The simulated particles ejected by the hypothetical parent bodies were treated in the same rigorous manner. Although we consider the similarity of the observed radiant and the dynamically modeled radiant as sufficient evidence for association with the hypothetical parent body when following the approach of \citet{vaubaillon2005new}, there are several points worth discussing.

All meteoroid orbits used in the analysis of \citet{segon2014parent} were generated using the UFOorbit software package by \citet{sonotaco2009meteor}. As this software does not estimate errors of the observations, or the errors of calculated orbital elements, it is not possible to consider the real precision of the individual meteoroid orbits used in the initial search analysis. Furthermore, all UFOorbit generated orbits are calculated on the basis of the meteor's average geocentric velocity, not taking the deceleration into consideration. This simplification introduces errors in orbital elements of very slow, very long, and/or very bright meteors. The real impact of this simplification is discussed in \citet{segon2014draconids} where the 2011 Draconid outburst is analyzed. Two average meteoroid orbits generated from average velocities were compared, one with and one without the linear deceleration model. These two orbits differed by as much as 0.06 in D$_{SH}$ (D$_{H}$ = 0.057, D$_{D}$ = 0.039). The deviation between the orbits does not necessarily mean that the clustering would not be determined, but it does mean that those orbits will certainly differ from the orbits generated with deceleration taken into account, as well as differing from the numerically generated orbits of hypothetical parent bodies. Consequently, the radiant locations of slower meteors can be, besides the natural radiant dispersion, additionally dispersed due to the varying influence of the deceleration in the position of the true radiant. This observation is not only relevant for UFOorbit, but for all software which potentially does not properly model the actual deceleration. CAMS Coincidence software uses an exponential deceleration model \citep{jenniskens2011cams}, however not all meteors decelerate exponentially as was shown in \citet{borovivcka2007atmospheric}. The real influence of deceleration in radiant dispersion will be a topic of some future work. Undoubtedly an important question is whether the dispersion caused by the improperly calculated deceleration of slow (e.g., generated by near-Earth objects) meteors can render members of a meteoroid stream to be unassociated with each other by the automated stream searching methods.

Besides the lack of error estimation of meteor observations, parent bodies on relatively unstable orbits are observed over a short observation arc, thus they often do not have very precise orbital element solutions. Moreover, unknown past parent body activity presents a seemingly unsolvable issue of how the parent orbit could have been perturbed on every perihelion pass close to the Sun. Also if the ejection modeling assumed that the particle ejection occurred during a perihelion passage when the parent body was not active, there would be no meteors present when the Earth passes through the point of the falsely predicted filament. On the other hand, if the Earth encounters meteoroids from a perihelion passage of particularly high activity, an unpredicted outburst can occur. \citet{vaubaillon2015latest} discuss the unknowns regarding parent bodies and the problems regarding meteor shower outburst prediction in greater detail.

Another fundamental problem that was encountered during this analysis was the lack of any rigorous definitions of what meteor showers or meteoroid streams actually are. Nor is there a common consensus to refer to. This issue was briefly discussed in \citet{brown2010meteoroid} and no real advances towards a clear definition have been made since. We can consider a meteor shower as a group of meteors which annually appear near the same radiant and which have approximately the same entry velocity. To better embrace the higher-dimensional nature of orbital parameters and time evolution versus a radiant that is fixed year after year, this should be extended to mean there exists a meteoroid stream with meteoroids distributed along and across the whole orbit with constraints dictated by dynamical evolution away from the mean orbit. By using the first definition however, some meteor showers caused by Jupiter-family comets will not be covered very well, as they are not active annually. The orbits of these kinds of meteor showers are not stable in the long term due to the gravitational and non-gravitational influences on the meteoroid stream.\footnote{\label{vaubaillonIMC2014}\url{http://www.imo.net/imc2014/imc2014-vaubaillon.pdf}.} On the other hand, if we are to consider any group of radiants which exhibit similar features but do not appear annually as a meteor shower, we can expect to have thousands of meteor shower candidates in the near future.

It is the opinion of the authors that with the rising number of observed multi-station video meteors, and consequently the rising number of estimated meteoroid orbits, the number of new potential meteor showers detected will increase as well, regardless of the stream search method used. As a consequence of the vague meteor shower definition, several methods of meteor shower identification have been used in recent papers. \citet{vida2014meteor} discussed a rudimentary method of visual identification combined with D-criterion shower candidate validation. \citet{rudawska2014new} used the Southworth and Hawking D-criterion as a measure of meteoroid orbit similarity in an automatic single-linkage grouping algorithm, while in the subsequent paper by  \citet{rudawska2015independent}, the geocentric parameters were evaluated as well. In \citet{jenniskens2016cams} the results of the automatic grouping by orbital parameters were disputed and a manual approach was proposed. 

Although there are concerns about automated stream identification methods, we believe it would be worthwhile to explore the possibility of using density-based clustering algorithms, such as DBSCAN or OPTICS algorithms by \citet{kriegel2005density}, for the purpose of meteor shower identification. They could possibly discriminate shower meteors from the background as they have a notion of noise and varying density of the data. We also strongly encourage attempts to define meteor showers in a more rigorous manner, or an introduction of an alternate term which would help to properly describe such a complex phenomenon. The authors believe that a clear definition would be of great help in determining whether a parent body actually produces meteoroids – at least until meteor observations become precise enough to determine the connection of a parent body to a single meteoroid orbit.

\section{Conclusion}
From this work, we can conclude that the following associations between newly discovered meteoroid streams and parent bodies are validated:
\begin{itemize}
        \item \#549 FAN 49 Andromedids and Comet 2001 W2 Batters
        \item \#533 JXA July $\xi$ Arietids and Comet C/1964 N1 Ikeya
        \item \#539 ACP $\alpha$ Cepheids and Comet P/255 Levy
        \item \#541 SSD 66 Draconids and Asteroid 2001 XQ
        \item \#751 KCE $\kappa$ Cepheids and Asteroid 2009 SG18
        \item \#753 NED November Draconids and Asteroid 2009 WN25
        \item \#754 POD $\psi$ Draconids and Asteroid 2008 GV
\end{itemize}
The connection between \#755 MID May $\iota$ Draconids and asteroid 2006 GY2 is not firmly established enough  and still requires some additional observational data before any conclusion can be drawn. The asteroidal associations are interesting in that each has a Tisserand parameter for Jupiter indicating it was a possible Jupiter-family comet in the past, and thus each may now be a dormant comet. Thus it may be worth looking for outgassing from asteroids 2001 XQ, 2009 SG18, 2009 WN25, 2008 GV, and even 2006 GY2 during their perihelion passages in the near future using high resolution imaging.

\begin{acknowledgements}
JV would like to acknowledge the availability of computing resources on the Occigen super computer at CINES (France) to perform the computations required for modeling the theoretical meteoroid streams. Special acknowledgement also goes to all members of the Croatian Meteor Network for their devoted work on this project. 
\end{acknowledgements}

\bibpunct{(}{)}{;}{a}{}{,} 

\bibliographystyle{aa}
\bibliography{Segon_2016A&A_bibliography}

\end{document}